\documentstyle[12pt]{article}

\begin{document}

\thispagestyle{empty}
\setcounter{page}{0}

\vspace*{-8ex}
 \null \hfill MPI-PhT/95-109 \\
 \null \hfill November 1995 \\[15mm]

\begin{center}

{\LARGE\bf
  q-deformed Phase Space and\\  [1.2ex]
  its Lattice Structure} \\  [11.5ex]

{\large M. Fichtm\"uller$^1$, A. Lorek\footnote{Supported by the
German-Israeli Foundation (G.I.F.)}$^{\,1}$ and J. Wess$^{1,2}$}
   \\ [7ex]

 $^1$ Max-Planck-Institut f\"{u}r Physik \\
   Werner-Heisenberg-Institut\\
 F\"{o}hringer Ring 6 , D - 80805 M\"{u}nchen, Germany \\
 Tel. (89) 32354-0, Fax (89) 3226704\\ [3ex]

 $^2$ Sektion Physik, Universit\"{a}t M\"{u}nchen \\
   Theresienstr.\,37, D - 80333 M\"{u}nchen, Germany \\ [18ex]

\end{center}

\begin{abstract}

\noindent
A q-deformed two-dimensional phase space is studied as a model for a
noncommutative phase space. A lattice structure arises that can be
interpreted as a spontaneous breaking of a continuous symmetry.
The eigenfunctions of a Hamiltonian that lives on such a lattice are
derived as wavefunctions in ordinary $x$-space.

\end{abstract}

\clearpage

\renewcommand{\theequation}{\arabic{section}.\arabic{equation}}
\newcommand{\nn}{\nonumber}
\newcommand{\be}{\begin{equation}}
\newcommand{\ee}{\end{equation}}
\newcommand{\bea}{\begin{eqnarray}}
\newcommand{\eea}{\end{eqnarray}}
\newcommand{\ZZ}{\mbox{Z\hspace{-1.35mm}Z}}

\section*{Introduction}

Quantum groups are a generalization of the concept of symmetries.
They act on noncommutative spaces that inherit a well-defined mathematical
structure from the quantum group symmetries.\\
Starting from such a noncommutative space as configuration space we
generalize it to a phase space where noncommutativity is already intrinsic
for a quantum mechanical system. The definition of this noncommutative
phase space is derived from the noncommutative differential structure on
the configuration space.\\
The thus obtained noncommutative phase space is a q-deformation of the
quantum mechanical phase space and we can apply all the machinery as we
have learned it from quantum mechanics. We shall associate essentially
selfajoint operators in a Hilbert space with observables and define the
time development by a Schroedinger equation.

As an interesting result we shall see that the observables that are
associated with position and momentum in a natural way have a discrete
spectrum. The q-deformed phase space puts physics on a q-lattice. This
lattice can be imbedded in an ordinary quantum mechanical phase space
and q-deformation has many similarities to a spontaneous breaking of a
symmetry.\\
The operators exhibiting the discrete spectrum will transform nonlinearly
under a continuous translation. For certain Hamiltonians the time
development of the system takes place entirely on the lattice. This shows
that dynamics can deform continuous space to a lattice structure.

In chapter 1 we define the concept of a q-deformed two-dimensional phase
space. In chapter 2 we study the Hilbert space representations of the
q-deformed phase space operators. This shows their discrete spectrum.
We also show that these operators can be represented in the usual
Hilbert space of nondeformed quantum mechanics \cite{lowe}.
In chapter 3 we show that these representations can be reduced to the
representations mentioned before.\\
In chapter 4 we show that translational invariance can be defined in the
thus obtained subspaces of the ordinary Hilbert space and that the
q-deformed variables transform nonlinearly.\\
In chapter 5 we finally study a simple quantum mechanical system formulated
in terms of q-deformed phase space variables and we find the eigenfunctions
of the corresponding Hamiltonian as wavefunctions in the usual quantum
mechanical formulation.\\

\section{The Algebra}

\setcounter{equation}{0}

The simplest q-deformed differential calculus in one dimension is based
on the following Leibniz rule \cite{wz}:
\be
\partial_{\rm x} \, {\rm x} = 1 + q\, {\rm x} \,\partial_{\rm x}
\label{11}
\ee
This could be the starting point for a q-deformed Heisenberg algebra.
However, if x is assumed to be a hermitean operator in a Hilbert space
the usual quantization rule $p\rightarrow -i\,\partial_{\rm x}$ does
not yield a hermitean momentum operator. This can be seen by comparing
(\ref{11}) with its conjugate relation:
\be
{\rm x}\, \overline{\partial}_{\rm x}
= 1 + q\, \overline{\partial}_{\rm x} \,{\rm x}
\hspace{5mm},\hspace{10mm} \overline{q} = q
\label{12}
\ee
Nevertheless we are going to define a conjugation operation on the algebra
x, $\partial_{\rm x}$ which is consistent with (\ref{12}). This is done
with the help of a scaling operator:
\be
\Lambda = 1 + (q-1)\,{\rm x} \,\partial_{\rm x}
\label{13}
\ee
{}From (\ref{11}) follows
\be
\Lambda\, {\rm x} = q\,{\rm x}\,\Lambda \;\;\;, \;\;\;\;\;
\Lambda\, \partial_{\rm x} = q^{-1}\,\partial_{\rm x}\,\Lambda
\label{14}
\ee
The occurence of such a scaling operator is typical for a noncommutative
differential calculus derived from quantum group symmetries.
For $q=1$ we would find $\Lambda=1$, in the undeformed case the scaling
operator $\Lambda$ is not at our disposal.

If we define conjugation by
\be
\overline{\partial}_{\rm x} = - q^{-1}\Lambda^{-1}\partial_{\rm x}
\;\;\; , \;\;\;\;\; \overline{\rm x} = {\rm x}
\label{15}
\ee
then the equations (\ref{11}) and (\ref{12}) are consistent. Conjugation,
being an involution, tells us that
\be
\partial_{\rm x} = - q^{-1}\overline{\partial}_{\rm x}\,\overline{\Lambda}^{-1}
\label{16}
\ee
This is only possible if
\be
\overline{\Lambda} = q^{-1}\,\Lambda^{-1}
\label{17}
\ee
which can be verified by a direct calculation.

The obvious choice for a hermitean momentum operator is
\be
P = - \frac{i}{2} \,( \partial_{\rm x} - \overline{\partial}_{\rm x})
\label{18}
\ee
where $\overline{\partial}_{\rm x}$ has to be taken from (\ref{15}).
With equation (\ref{13}) this is an expression in terms of x and
$\partial_{\rm x}$.

Commuting $P$ through x yields:
\bea
P\,{\rm x} &=& -\frac{i}{2}\,(1+\frac{1}{q}) - \frac{i}{2}\,{\rm x}\,
(q\,\partial_{\rm x} - \frac{1}{q}\,\overline{\partial}_{\rm x}) \nn\\
  &=& -\frac{i}{2}\,(1+\frac{1}{q}) + q\,{\rm x}\,P
       - \frac{i}{2}\,(q-\frac{1}{q})\,{\rm x} \,\overline{\partial}_{\rm x}
\label{19}
\eea
We find
\be
P\,{\rm x} - q\,{\rm x}\,P = -\frac{i}{2}\,(1+\frac{1}{q})\,
\left[ 1+ (q-1)\,{\rm x}\,\overline{\partial}_{\rm x} \right]
\label{110}
\ee
For hermitean $P$ and x the right hand side of the q-deformed commutator had
to be an operator. Any such operator might be expressed in terms of the
product of a unitary and a hermitean operator.
Using (\ref{12}) and (\ref{13}) we see that the right hand side of (\ref{110})
can be expressed in terms of $\overline{\Lambda}$.\\[2mm]
With a simple redefinition
\be
U = q^{\frac{1}{2}}\,\overline{\Lambda} \;\;\; , \;\;\;\;\;
X = \frac{1+q}{2q}\;{\rm x}
\label{111}
\ee
we arrive at a q-deformed Heisenberg relation:\\
\be
\begin{array}{c}
q^{\frac{1}{2}}\,X\,P - q^{-\frac{1}{2}}\,P\,X = i\,U \\[4mm]
U\,X = q^{-1}\,X\,U \;\;\; , \;\;\;\;\; U\,P = q\,P\,U
\end{array}
\label{112}
\ee
with conjugation properties
\be
\overline{P} = P \;\; , \;\;\; \overline{X} = X \;\; , \;\;\;
\overline{U} = U^{-1}
\label{113}
\ee
This will be the starting point of our investigations.

A q-deformed quantization forced us to introduce an additional unitary
operator $U$, very much in the same way as ordinary quantization forces
us to a purely imaginary right hand side of the commutator.\\

\section{The Representations}

\setcounter{equation}{0}

Representations of the algebra (\ref{112}) have been constructed in
references \cite{schw}, \cite{heb}, \cite{sch}. We are shortly listing
the results.\\
As we are interested in representations where $P$ and $X$ are
represented by essentially selfajoint operators (so that they can be
diagonalized) we may start as well with a representation where $P$ is
diagonal. We note that a rescaling of $P$ and $X$ by a real parameter $s$
($P \rightarrow s P$, $X \rightarrow s^{-1}X$) does not change the algebra
nor the conjugation properties. Therefore we can always scale a
nonvanishing eigenvalue to $+1$.
It follows from the commutation property of $P$ and $U$, that such a
representation will have all the eigenvalues $q^n$ ($n\in \ZZ$). The
corresponding eigenstates form a basis of a representation space for the full
algebra. Analogously we could have obtained a representation with
eigenvalues $-q^n$.\\
These representations we denote by $\hat{P}$, $\hat{X}$, $\hat{U}$, the
corresponding eigenstates by $|n,\sigma\!>$, $\sigma\!=\!+,-$
respectively. It is easy to check that these representations are of the
following form:
\bea
&&\hat{P}\,|n,\sigma>\; =\; \sigma \,q^n\,|n,\sigma> \nn\\
&&\hat{X}\,|n,\sigma>\; =\; i\sigma \frac{q^{-n}}{q-q^{-1}}
                       \left( q^{\frac{1}{2}} |n-1,\sigma>
                       -\, q^{-\frac{1}{2}} |n+1,\sigma> \right)\nn\\
&&\hat{U}\,|n,\sigma>\; =\; |n-1,\sigma>\label{21}\\[2mm]
&&<n',\sigma'\,|\,n,\sigma>\; =\; \delta_{nn'}\,\delta_{\sigma\sigma'}\nn
\eea
For each choice of $\sigma$, this forms a representation. However for a
fixed value of $\sigma$, $\hat{X}$ is not essentially selfadjoint.
There is a one parameter family of selfadjoint extensions - none of
which satisfies the algebra. It is, however, possible to find a
selfadjoint extension that satisfies the algebra in a representation
where $\sigma$ takes both values. This will be our representation space
in what follows. When diagonalized, $\hat{X}$ will have the eigenvalues
$\mp \lambda^{-1} q^{-\frac{1}{2}}\,q^{\nu}\,\;(\nu\in\ZZ)$.
With the help of q-deformed cosine and sine functions, the change of
basis can be achieved.

The elements of the q-deformed algebra (\ref{112}) can also be expressed
in terms of the operators of an undeformed algebra, which we denote by
$p$, $x$; they satisfy $[x,p]=i$. (This operator $x$ should not be
confused with the element x in chapter 1.)\\
The relevant relations are:
\bea
P &=& p \nn\\
X &=& \frac{[z+\frac{1}{2}]}{z+\frac{1}{2}}\, x \label{22}\\
U &=& q^z \nn
\eea
where
\bea
z  &=& -\frac{i}{2}\,\left( xp+px \right) \label{23}\\[2mm]
{[A]} &=& \frac{q^A - q^{-A}}{q-q^{-1}}\hspace{20mm}\mbox{for any A} \nn
\eea
The algebraic relations (\ref{112}) follow from the canonical commutation
properties of $p$, $x$. We outline the proof \cite{curt}. We start from
\be
x\,z = (z+1)\, x \;\;\; , \;\;\;\; p\, z = (z-1)\,p
\label{24}
\ee
so that
\be
x\,f(z)=f(z+1)\, x \;\;\; , \;\;\;\; p\, f(z) = f(z-1)\,p
\label{25}
\ee
for functions $f(z)$. We also use the identity
$z=-ipx+\frac{1}{2}=-ixp-\frac{1}{2}$.
Inserting (\ref{22}) into (\ref{112}) yields
\be
q^{\frac{1}{2}} \,\frac{[z+\frac{1}{2}]}{z+\frac{1}{2}}\,xp
  - q^{-\frac{1}{2}} \,\frac{[z-\frac{1}{2}]}{z-\frac{1}{2}}\,px\nn\\
= i q^{\frac{1}{2}}\,[z+\frac{1}{2}] -
    i q^{-\frac{1}{2}}\, [z-\frac{1}{2}] \; = \; iq^z
\label{26}
\ee
The conjugation properties (\ref{113}) follow from the hermiticity of $x$
and $p$.\\[2mm]
Algebraically the form of the relations (\ref{22}) can be changed by
canonical transformations on $x$ and $p$.
An interesting class of such canonical transformations is:
\be
\tilde{p} = f(z)\,p \;\; ,\;\;\; \tilde{x} = x\,f^{-1}(z)
\label{27}
\ee
The hermiticity of $\tilde{p}$ and $\tilde{x}$ demands:
\be
\overline{f}(\overline{z}) = f(z+1)
\label{28}
\ee
This condition is e.g. satisfied for
\be
f^{-1}(z) = \frac{[z-\frac{1}{2}]}{z-\frac{1}{2}}
\label{29}
\ee
With this choice for a canonical transformation the relations (\ref{22})
become:
\bea
P &=& \frac{[\tilde{z}-\frac{1}{2}]}{\tilde{z}-\frac{1}{2}}\,\tilde{p}
\nn\\
X &=& \tilde{x} \label{210}\\
U &=& q^{\tilde{z}} \nn
\eea
The standard Hilbert space representation of $p$, $x$ leads to a
representation of $P$, $X$ and $U$ via the relation (\ref{22}) or
(\ref{210}). How this representation is related to the representations
(\ref{21}) will be discussed in the next chapter.\\
In this context we will encounter rescaled eigenvalues of $P$,
$\sigma s q^n$ ($1\!\!\leq\!\! s \!\!<\!\! q$). To distinguish such
representations we introduce an operator $S$ which commutes with $P$, $X$
and $U$, which is hermitean and has a spectrum ranging from $1$ to $q$.
\be
\begin{array}{l}
S^+ = S\;\; ,\;\; [S,P] = [S,X] = [S,U] = 0 \\[2mm]
S\,|s> \;=\; s\,|s> \;\;\;\;\;\;1\leq s < q \\[2mm]
<s'\,|\,s>\;=\; \delta(s'-s)
\end{array}
\label{211}
\ee
With $\hat{X}$, $\hat{P}$, $\hat{U}$ we will find another representation
of (\ref{112}):
\bea
P &=& S\,\hat{P} \nn\\
X &=& S^{-1}\,\hat{X} \label{212}\\
U &=& \hat{U} \nn
\eea
The eigenstates of $P$ are $|n,\sigma>|s>$ with the eigenvalue
$\sigma  s q^n$.

\section{Reduction of Representations}

\setcounter{equation}{0}

Relations (\ref{22}) allow to represent $P$, $X$, $U$ in the
Hilbert space of ordinary quantum mechanics where we choose a
momentum representation
\be
\begin{array}{l}
p\,|p_0> \;=\, p_0 \,|p_0> \\[2mm]
<p_0'\,|\,p_0>\; =\, \delta(p_0'-p_0)
\end{array}
\label{31}
\ee
Equation (\ref{21}) suggests to change the above basis to the new
basis:
\be
|n,\sigma>|s>\; = \int dp_0 \,q^{\frac{n}{2}}\,\delta(p_0-\sigma s q^n)\;
|p_0>
\label{32}
\ee
It is easy to check that with the normalization (\ref{31}) the states
$|n,\sigma>$ and $|s>$ are normalized as in (\ref{21}) and (\ref{211}).
The inverse transformation is:
\be
|p_0>\; = \int_{1}^{q} ds\, \sum_{n=-\infty}^{\infty}\, \sum_{\sigma=+,-}
\, q^{\frac{n}{2}}\, \delta(p_0-\sigma s q^n)\; |n,\sigma>|s>
\label{33}
\ee
We now show that this change of basis reduces the representation of $P$,
$X$, $U$ in terms of $x$ and $p$ to representations as we encountered them
in (\ref{21}).\\
For the operator $P$, this is trivial:
\be
P\, |n,\sigma>|s>\; =\; \sigma s q^n \,|n,\sigma>|s>
\label{34}
\ee
For the operators $U$ and $X$ it requires a short calculation. We know
that $x$ is represented by $i\frac{\partial}{\partial p}$ when acting
on a wave function. Therefore:
\be
U\,f(p) = q^{\frac{1}{2}}q^{p\frac{\partial}{\partial p}}\,f(p) =
  q^{\frac{1}{2}}f(qp)
\label{35}
\ee
This has as an immediate consequence that
\be
U\,|n,\sigma>|s>\; =\; |n-1,\sigma>|s>
\label{36}
\ee
The operator $X$ of equation (\ref{22}) can be rewritten as follows:
\be
X = \frac{i}{q-q^{-1}}\,\left( q^{\frac{1}{2}}\,U -
    q^{-\frac{1}{2}}\,U^{-1} \right) \; P^{-1}
\label{37}
\ee
This again implies (\ref{21}) directly. The states $|n,\sigma>|s>$ carry
a representation of $P$, $X$ and $U$ for a fixed value of $s$.

Relations (\ref{22}) can be inverted formally:
\bea
p &=& P \nn\\
z &=& h^{-1}\,{\rm ln}\,U \label{38}\\
x &=& \frac{i}{h}\,{\rm ln}\left( q^{\frac{1}{2}}U \right) \;P^{-1}
\hspace{15mm}, \;\;\;\;\;\; q=e^h \nn
\eea
Hermiticity properties of $x$ and $z$ may be used to make ${\rm ln}\,U$
less ambiguous.\\
Of course we cannot succeed in constructing selfadjoint representations
of the canonical operators $x$ and $p$ in the representation space of
(\ref{21}). This manifests itself when we apply ${\rm ln}\,U$ to any of the
states $|n,\sigma>$, we get an expansion of the logarithm around zero or
infinity.\\[-5mm]

There are states, not normalizable, that are eigenstates of the operator
$U$:
\bea
&&|\phi_0>\; =\; \frac{1}{\sqrt{2\pi}} \,\int_{1}^{q} ds\, \sum_{n,\sigma}
\, q^{\frac{n}{2}} \, |n,\sigma>|s>
= \frac{1}{\sqrt{2\pi}} \, \int dp_0\,|p_0>\nn\\
&&U|\phi_0>\; =\; q^{\frac{1}{2}}\,|\phi_0>
\label{39}
\eea
and
\bea
|\phi_m> \;=\; P^m\,|\phi_0>\nn\\
U|\phi_m>\; =\; q^{m+\frac{1}{2}}\,|\phi_m>
\label{310}
\eea
For these states, in a formal way, we find:
\be
x \,|\phi_m>\; =\; i m\:|\phi_{m-1}>\hspace{20mm}m\geq0
\label{311}
\ee
Eigenstates of $x$ can be obtained from the states $|\phi_m>$:
\bea
&&|x_0> \;=\; \sum_{m=0}^{\infty}\;\frac{(-i)^m\,x_0^m}{m!} \;|\phi_m>\nn\\
&&x\,|x_0> \;=\; x_0\;|x_0>
\label{312}
\eea
After these heuristic arguments we show that the states $|x_0>$ are well
defined. For this purpose it is convenient to use (\ref{33}) and write
$|x_0>$ in the following form:
\bea
|x_0> &=&\frac{1}{\sqrt{2\pi}}\,\sum_{m=0}^{\infty}\;\frac{(-i)^m\,x_0^m}{m!}
\;P^m\;\int dp_0\;\int_{1}^{q} ds\, \sum_{n,\sigma}\,\delta(p_0-\sigma s q^n)\,
\, q^{\frac{n}{2}} \, |n,\sigma>|s>\nn\\
&=& \frac{1}{\sqrt{2\pi}}\,\int dp_0\;\int_{1}^{q} ds\, \sum_{n,\sigma}\,
e^{-ix_0\,p_0}\,\delta(p_0-\sigma s q^n)\,\, q^{\frac{n}{2}} \, |n,\sigma>|s>
\label{313}
\eea
This is nothing but the Fourier transform of the states $|p_0>$. The
relation
\be
<x_0'\,|\,x_0> \;=\; \delta(x_0'-x_0)
\label{314}
\ee
follows after a short calculation from the normalization of the states
$|n,\sigma>|s>$.\\[2mm]
We have obtained a representation of $x$ that is essentially selfajoint
from a reducible representation of $X$ and $P$.\\

\section{Translational Invariance}

\setcounter{equation}{0}

The unitary operator that represents translation is $e^{ipa}$. From
(\ref{38}) follows that the representation space of $P$, $X$ and $U$
is left invariant under such a translation. This shows that there is a
map of the operators $P$, $X$ and $U$ into operators $P'$, $X'$, $U'$
that corresponds to a translation of the system:
\bea
P' &=& e^{iap}\;P\;e^{-iap} \;=\; P \nn\\
U' &=& e^{iap}\;U\;e^{-iap} \;=\; U\,e^{iaq^{-1}(1-q)P} \label{41}\\
X' &=& e^{iap}\;X\;e^{-iap} \;=\; \frac{i}{\lambda}\left(
   q^{\frac{1}{2}}\,U\,e^{iaq^{-1}(1-q)P}
         - q^{-\frac{1}{2}}\,U^{-1}\,e^{ia(q-1)P}\right)\;P^{-1} \nn
\eea
This exemplifies an idea of Snyder who implemented Lorentz invariance
in quantized spaces \cite{sny}. Formula (\ref{41}) resembles formulas
for nonlinear transformation laws:
\be
\delta X = X' - X = q^{-\frac{1}{2}}\,(q+1)^{-1}
                   \left( U+U^{-1} \right)\,a
\label{42}
\ee
For nonlinear transformation laws a shift in the origin is a
characteristic feature. Here $X$ is shifted by an operator.

Let us demonstrate the action of the translation on a state $|x_0>$
(\ref{313}) by acting on the Hilbert space of (\ref{21}):
\bea
e^{iap} \,|x_0> &=& \frac{1}{\sqrt{2\pi}}\,\int dp_0\;\int_{1}^{q} ds\,
\sum_{n,\sigma}\,e^{-ix_0\,p_0}\,\delta(p_0-\sigma s q^n)\,\,
q^{\frac{n}{2}} \, e^{iaP}\,|n,\sigma>|s>\nn\\
&=&  \frac{1}{\sqrt{2\pi}}\,\int dp_0\;\int_{1}^{q} ds\,
\sum_{n,\sigma}\,e^{-ix_0\,p_0}\,\delta(p_0-\sigma s q^n)\,\,
q^{\frac{n}{2}} \, e^{ia\sigma s q^n}\,|n,\sigma>|s>\nn\\
&=&  \frac{1}{\sqrt{2\pi}}\,\int dp_0\;\int_{1}^{q} ds\,
\sum_{n,\sigma}\,e^{-i(x_0-a)\,p_0}\,\delta(p_0-\sigma s q^n)\,\,
q^{\frac{n}{2}} \,|n,\sigma>|s>
\label{43} \\[1mm]
&=& |x_0-a>\nn
\eea
This clearly shows that the translation acts only on the lattice part
of the decomposition (\ref{212}) and not on the operator $S$ - again
a situation we are used to from realizing spontaneous symmetry breaking
by nonlinearly transforming fields.\\

\section{q-deformed Dynamics}

\setcounter{equation}{0}

The simplest q-deformed Hamiltonian is
\be
H = \frac{1}{2}\,P^2
\label{51}
\ee
We know that this Hamiltonian has eigenvalues $\frac{1}{2}s^2 q^{2n}$.
If we associate canonical variables as in (\ref{210}) (in the following
we drop the tilde on $x$ and $p$) this describes an interacting system:
\be
H = \frac{2}{\lambda^2}\;\;p\;
      \frac{q+q^{-1} - 2\,{\rm cos}((xp+px)h)}{1+(xp+px)^2}\;p
\label{52}
\ee
We construct the eigenfunctions of this Hamiltonian in the
$x$-representation. Through the identification (\ref{210}) it is
easy to find the eigenfunctions of the operator $X$
\be
|\nu,\tau>|s> \,= \int dx_0 \;\lambda^{-\frac{1}{2}}\,
        q^{\frac{\nu}{2}-\frac{1}{4}}\,
        \delta(s\,x_0-\lambda^{-1}\tau q^{\nu-\frac{1}{2}})\;|x_0>
\label{53}
\ee
The normalization has been chosen such that
\bea
<x_0'\,|\,x_0> &=& \delta(x_0'-x_0)\nn\\[1mm]
<\nu',\tau'\,|\,\nu,\tau> &=& \delta_{\nu'\nu}\;\delta_{\tau'\tau}
\label{54}\\[1mm]
<s'\,|\,s> &=& \delta(s'-s)\nn
\eea
and we have eigenstates of the operator $X$ with eigenvalues
$\tau q^{-\frac{1}{2}}\lambda^{-1}\,s^{-1}\,q^{\nu}$ as we
expect them from the representation theory of (\ref{112}).\\
In reference \cite{heb} we have learned how to transform such a basis
into a basis of momentum eigenstates. Such a transformation will
diagonalize the Hamiltonian. We choose the normalization of the $X$
eigenstates consistent with the representation (\ref{210}) of $U$.\\
The momentum eigenstates are:
\bea
|2m,\sigma> &=& \frac{1}{2}\,N_q \sum_{\nu} q^{\nu+m}\left[
 \cos_q(q^{(\nu+m)}) \left(|2\nu,\tau\!\!=\!\!+\!\!>
+\;|2\nu,\tau\!\!=\!\!-\!\!> \right) \right. \label{55}\\
 &&\hspace*{20mm} \left. -i\sigma \;\sin_q(q^{2(\nu+m)})
   \left(|2\nu\!\!+\!\!1,\tau\!\!=\!\!+\!\!>-\;
|2\nu\!\!+\!\!1,\tau\!\!=\!\!-\!\!>\right) \right]\nn\\[3mm]
|2m\!\!+\!\!1,\sigma> &=& \frac{1}{2}\,N_q \sum_{\nu} q^{\nu+m}\left[
 \cos_q(q^{(\nu+m)}) \left(|2\nu\!\!-\!\!1,\tau\!\!=\!\!+\!\!>+\;
|2\nu\!\!-\!\!1,\tau\!\!=\!\!-\!\!>\right)\right.\nn\\
 &&\hspace*{20mm} \left. -i\sigma \,\sin_q(q^{2(\nu+m)})
   \left(|2\nu,\tau\!\!=\!\!+\!\!>-\;|2\nu,\tau\!\!=\!\!-\!\!>\right)\right]\nn
\eea
The functions $\cos_q$ and $\sin_q$ were defined by Koornwinder and
Swarttouw \cite{koorn} with the following change of notation:
$\cos_q(z) = \cos(z;q^{-4})$, $\sin_q(z) = \sin(z;q^{-4})$.\\
Of importance is the following relation
\bea
&\frac{1}{z} \left[ \cos_q(z) - \cos_q(q^{-2}z) \right] &=
-q^{-2}\,\sin_q(q^{-2}z)\nn\\[2mm]
&\frac{1}{z} \left[ \sin_q(z) - \sin_q(q^{-2}z) \right] &= \cos_q(q^{-2}z)
\label{56}
\eea
When we insert the wavefunction (\ref{53}) into (\ref{55}) we get the
wavefunctions for the momentum eigenstates.\\
For the eigenstate of the Hamiltonian we consider the state
\bea
&&\frac{1}{\sqrt{2}}\,\left( \, |2m,\sigma\!\! = \!\!+\!\!>|s> + \;
|2m,\sigma\!\! = \!\!-\!\!>|s>\, \right) \label{57}\\
&&= \frac{1}{\sqrt{2}} N_q \sum_{\nu,\tau} q^{\nu+m} \cos_q(q^{2(\nu+m)})
  \int dx_0\, \delta(s x_0-\tau\lambda^{-1}q^{2\nu-\frac{1}{2}})
  \lambda^{-\frac{1}{2}}\,q^{\nu-\frac{1}{4}}\;|x_0>\nn
\eea
The corresponding eigenfunction of the Hamiltonian (\ref{52}) to the
eigenvalue $\frac{1}{2}s^2q^{4m}$ is:
\be
\Psi_{2m,s}(x) = \sqrt{\frac{\lambda}{2}}\,N_q\,\sum_{\nu,\tau}
\tau s x \,q^{m+\frac{1}{4}}\,\cos_q(q^{2m+\frac{1}{2}} \lambda\tau s x)
\,\delta(s x -\tau\lambda^{-1}q^{2\nu-\frac{1}{2}})
\label{58}
\ee
Forgetting the way we obtained this wavefunction we can verify that
it is indeed an eigenfunction of the Hamiltonian (\ref{52}) by
making use of (\ref{56}). The Hamiltonian when applied to a function
acts in the following way
\be
H f(x) = -\frac{1}{2\lambda^2}\,\frac{1}{x^2}
           \left( q\,f(\frac{1}{q^2}x) - (q+\frac{1}{q})\,f(x)
           + \frac{1}{q}\,f(q^2x) \right)
\label{59}
\ee
A short calculation verifies:
\be
H\;\Psi_{2m,s} = \frac{s^2}{2}\,q^{4m}\;\Psi_{2m,s}
\label{510}
\ee
The eigenfunctions are normalized to a $\delta$-function in the
$s$-space. In a similar way we can treat all the eigenfunctions
of $H$.

It is interesting to note that the Hamiltonian (\ref{52}) forces the
system to live on a lattice. If there is an interaction which is a
polynomial in $P$ and $X$ the system will never leave this lattice
because the eigenvalue $s$ of $S$ will not change.


\begin{thebibliography}{50}

\bibitem{lowe} A. Lorek, J. Wess:\\
               Dynamical Symmetries in q-deformed Quantum Mechanics,\\
               Z. Phys. C 67 (1995) 671

\bibitem{wz} J. Wess, B. Zumino:\\
             Covariant Differential Calculus on the Quantum Hyperplane,\\
             Nucl. Phys. B (Proc. Suppl.) 18B (1990) 302

\bibitem{schw} J. Schwenk, J. Wess:\\
               A q-deformed Quantum Mechanical Toy Model,\\
               Phys. Lett. B 291 (1992) 273

\bibitem{heb} A. Hebecker, S. Schreckenberg, J. Schwenk, W. Weich, J. Wess:\\
              Representations of a q-deformed Heisenberg Algebra,\\
              Z. Phys. C 64 (1994) 355

\bibitem{sch} J. Schwenk:\\
              q-deformed Fourier Theory,\\
              preprint MPI-PhT/94-36, hep-th/9406168

\bibitem{curt} T. Curtright, C. Zachos:\\
               Deforming Maps for Quantum Algebras,\\
               Phys. Lett. B 243 (1990) 237

\bibitem{sny} H. S. Snyder:\\
              Quantized Space-Time,\\
              Phys. Rev. 71 (1947) 38

\bibitem{koorn} T. H.  Koornwinder, R. F. Swarttouw:\\
                On q-analogues of the Fourier and Hankel Transforms,\\
                Trans. AMS 333 (1992) 445

\end{thebibliography}
\end{document}